\documentclass[twocolumn,showpacs,superscriptaddress,amsmath,amssymb]{revtex4-1}
\usepackage{graphicx}
\usepackage{epsfig}
\usepackage{dcolumn}
\usepackage{multirow}
\usepackage{bm}
\usepackage{overpic}
\usepackage{color}
\usepackage{lineno}
\usepackage{setspace}
\usepackage{url}
\usepackage{subfigure}
\include{def-com}

\def \jpsi {J/\psi}
\def \epem {e^+e^-}

\def \pzpz {\pi^0\pi^0}
\def \pppm {\pi^{+}\pi^{-}}
\def \kpkm {K^{+}K^{-}}
\def \ks   {K^{*}(892)}
\def \ksks {K^{*+}(892)K^{*-}(892)}

\def \piz  {\pi^0}

\def \gev  {\mbox{GeV}}

\def \gevcc{\mbox{GeV/$c^2$}}
\def \mev  {\mbox{MeV}}

\def \mevcc{\mbox{MeV/$c^2$}}
\def \ipb  {\mbox{pb$^{-1}$}}
\def \ifb  {\mbox{fb$^{-1}$}}
\begin{document}
\title{\boldmath Observation of a resonant structure in $\epem \to \kpkm\pzpz$}
\author{
\begin{small}
M.~Ablikim$^{1}$, M.~N.~Achasov$^{10,e}$, P.~Adlarson$^{63}$,
S. ~Ahmed$^{15}$, M.~Albrecht$^{4}$, A.~Amoroso$^{62A,62C}$,
Q.~An$^{59,47}$, ~Anita$^{21}$, Y.~Bai$^{46}$, O.~Bakina$^{28}$
% ...
\\
\vspace{0.2cm}
(BESIII Collaboration)\\
\vspace{0.2cm} {\it
$^{1}$ Institute of High Energy Physics, Beijing 100049, People's Republic of China\\
% ...
$^{4}$ Bochum Ruhr-University, D-44780 Bochum, Germany\\
% ...
$^{10}$ G.I. Budker Institute of Nuclear Physics SB RAS (BINP), Novosibirsk 630090, Russia\\
% ...
$^{15}$ Helmholtz Institute Mainz, Johann-Joachim-Becher-Weg 45, D-55099 Mainz, Germany\\
% ...
$^{21}$ Indian Institute of Technology Madras, Chennai 600036, India\\
% ...
$^{28}$ Joint Institute for Nuclear Research, 141980 Dubna, Moscow region, Russia\\
% ...
$^{46}$ Southeast University, Nanjing 211100, People's Republic of China\\
$^{47}$ State Key Laboratory of Particle Detection and Electronics, Beijing 100049, Hefei 230026, People's Republic of China\\
% ...
$^{59}$ University of Science and Technology of China, Hefei 230026, People's Republic of China\\
% ...
$^{62}$ (A)University of Turin, I-10125, Turin, Italy; (B)University of Eastern Piedmont, I-15121, Alessandria, Italy; (C)INFN, I-10125, Turin, Italy\\
$^{63}$ Uppsala University, Box 516, SE-75120 Uppsala, Sweden\\
% ...
$^{e}$ Also at the Novosibirsk State University, Novosibirsk, 630090, Russia\\
% ...
}
\vspace{0.4cm}
}
\noaffiliation{}
\date{\today}
\begin{abstract}
Using $\epem$ collision data samples with center-of-mass energies ranging from 2.000 to 2.644~$\gev$, collected by the BESIII detector at the BEPCII collider, and with a total integrated luminosity of 300~$\ipb$, a partial-wave analysis is performed for the process $\epem \to \kpkm\pzpz$.
The total Born cross sections for the process $\epem \to$ $\kpkm\pzpz$, as well as the Born cross sections for the subprocesses $\epem\to\phi \pzpz$, $K^{+}(1460)K^{-}$, $K^{+}_{1}(1400)K^{-}$, $K^{+}_{1}(1270)K^{-}$ and $\ksks$, are measured versus the center-of-mass energy.
The corresponding results for $\epem \to \kpkm\pzpz$ and $\phi \pzpz$ are consistent with those of BaBar and have much improved precision.
By analyzing the cross sections for the four subprocesses, $K^{+}(1460)K^{-}$, $K^{+}_{1}(1400)K^{-}$, $K^{+}_{1}(1270)K^{-}$ and $\ksks$, a structure with mass $M$ = (2126.5 $\pm$ 16.8 $\pm$ 12.4)~$\mevcc$ and width $\Gamma$ = (106.9 $\pm$ 32.1 $\pm$ 28.1)~$\mev$ is observed with an overall statistical significance of 6.3$\sigma$, although with very limited significance in the subprocesses $\epem\to K^{+}_{1}(1270)K^{-}$ and $\ksks$.
The resonant parameters of the observed structure suggest it can be identified with the $\phi(2170)$,
thus the results provide valuable input to the internal nature of the $\phi(2170)$.
\end{abstract}

\maketitle

The vector meson state $Y(2175)$, denoted as the $\phi(2170)$ by the Particle Data Group (PDG)~\cite{PDG}, is currently one of the most interesting particles in light hadron spectroscopy.
The $\phi(2170)$ was first observed by  BaBar~\cite{Y2170Babar} and subsequently confirmed by
several other experiments~\cite{Y2170Babar1,Y2170belle,Y2170Bes,Y2170Bes3,2017bes32}.
The internal constituents of the $\phi(2170)$ are still unknown, which has stimulated extensive theoretical discussions.
Possible interpretations of the $\phi(2170)$ include a conventional $3^{3}S_{1}$ or $2^{3}D_{1}$ $s\bar{s}$ state~\cite{strange,2017ding,2017wang,2017afonin},
an $s\bar{s}g$ hybrid~\cite{2017ding, 2017ding2,hybrid2}, a tetraquark state~\cite{2017wang2,2017chen,2019ke,2017drenska},
a $\Lambda \bar{\Lambda}(^3S_1)$ bound state~\cite{2017zhao,2017deng,2017dong}, or a $\phi KK$ resonance state~\cite{2017oset}, etc., but no interpretation has yet been established.
Each of these theoretical models can accommodate a resonant state with parameters similar to those of the $\phi(2170)$, but they predict significantly different partial widths for individual decay modes, especially the $K^{(*)}K^{(*)}$ decay modes, where the $K^{(*)}$ is the ground or excited state of a $K$ meson with different spin-parities.
Consequently, studying the decay modes of the $\phi(2170)$, and precisely measuring their partial widths, plays a key role in determining the internal structure of the $\phi(2170)$.

The BESII collaboration searched for the decay $\phi(2170) \to K^{*0}(892)\bar{K}^{*0}(892)$ via $\jpsi\to\eta \phi(2170)$ by using 58~million $\jpsi$ events~\cite{ksksupper}. No significant signal was observed.
The BaBar collaboration performed an analysis of $\epem \to \kpkm\pppm$ and $\kpkm\pzpz$ using 454~$\ifb$ data via the initial state radiation (ISR) process~\cite{Y2170Babar}.
Beside clearly observing the process $\epem\to\phi \pi\pi$, abundant $K^*$ structures were observed in the $K\pi(\pi)$ invariant mass spectrum, such as the $K^*(892)$ and $K^*_2(1430)$, as well as the $K_1(1270)$ and $K_1(1400)$.
It is worth noting that only about 1\% of the
$\epem \to \kpkm\pppm$ events were from the subprocess $\epem\to K^{*0}(892)\bar{K}^{*0}(892)$, while roughly 30\% of the $\epem \to \kpkm\pzpz$ events were from $\epem\to K^{*+}(892)K^{*-}(892)$.
A comprehensive analysis, \emph{e.g.} a partial-wave analysis (PWA), is desired to resolve the contribution of individual components in these decays.

Besides an excited $\phi$ state, the quark model also predicts excited $\rho$ and $\omega$ states in the 2~$\gevcc$ mass range~\cite{Godfrey:1985xj}.
Finding this set of excited vector mesons would help establish the corresponding $\rho$, $\omega$, and $\phi$ meson families and would set a baseline for theoretical models.
Since these excited vector mesons can each decay into $K^{(*)}K^{(*)}$ final states, analyzing the $K^{(*)}K^{(*)}$ invariant mass spectra in $e^+e^-$ annihilation becomes an effective means to discover them.

In this Letter, we present a PWA of the process $\epem \to \kpkm\pzpz$ using data collected with the BESIII detector.
The ten data samples used in this analysis have center-of-mass (c.m.) energies ranging from 2.000 to 2.644~$\gev$ and have a total integrated luminosity of 300~$\ipb$.
The c.m. energy values and integrated luminosities of each data set are presented in Table I in the supplemental material~\cite{supply}.
Charge-conjugated processes are always included by default.

 Detailed descriptions of the design and performance of the BESIII detector can be found in Ref.~\cite{besint}.
 A Monte Carlo (MC) simulation based on {\sc Geant4}~\cite{geant4}, including the geometric description of the BESIII detector and its response, is used to optimize the event selection criteria, estimate backgrounds, and determine the detection efficiency.
 The signal MC samples are generated using the package {\sc ConExc}~\cite{conexc}, which incorporates a higher-order ISR correction.
 Background samples of the processes $\epem \to \epem$, $\mu^{+}\mu^{-}$ and $\gamma\gamma$ are generated with the {\sc Babayaga}~\cite{babayaga} generator, while $\epem \to$ hadrons and two photon events are generated by the {\sc Luarlw}~\cite{lumarlw} and {\sc Bestwogam}~\cite{bestwogam} generators, respectively.

 The selection criteria for charged tracks, particle identification (PID), and photon candidates are the same as those in Ref.~\cite{selec}.

 The process $\epem \to \kpkm\pzpz$ results in the final state $\kpkm\gamma\gamma\gamma\gamma$. Thus, candidate events with only
 two oppositely-charged kaons and at least four photons are selected.
 To improve the kinematic resolution and suppress background, a six-constraint (6C) kinematic fit imposing energy-momentum conservation, as well as two additional $\piz$ mass constraints, is carried out under the hypothesis $\epem\to\kpkm\piz\piz$.
The combination with minimum $\chi^ {2}_{6C}$ is retained for further analysis.
 The candidate events are required to satisfy $\chi^{2}_{6C}$~$<$ $80$.
 After the above selection criteria, detailed studies indicate that the backgrounds are negligible.

 Using the GPUPWA framework~\cite{PWAframe}, a PWA is performed on the surviving candidate events to disentangle the intermediate processes present in $\epem \to \kpkm\pzpz$.
 The quasi two-body decay amplitudes in the sequential decays are constructed using covariant tensor amplitudes~\cite{PWAtensor}. 
The intermediate states are parameterized with relativistic Breit-Wigner (BW) functions, except for the $f_{0}(980)$, which is described with a Flatt\'{e} formula~\cite{PWA:sigma1}.
 The resonance parameters of the $f_{0}(980)$ and the wide resonance $\sigma$ in the fit are fixed to those in Ref.~\cite{PWA:sigma1} and Ref.~\cite{PWA:sigma1,PWA:sigma2}, respectively, and those of other intermediate states are fixed to PDG values, or measured in the analysis.
To include the resolution for the narrow $\phi(1020)$ resonance, a Gaussian function is convolved with the BW function, but this is not done for the other resonances.
 The relative magnitudes and phases of the individual intermediate processes are determined by performing an unbinned maximum likelihood fit using MINUIT~\cite{PWA:minuit1}.

We start the fit procedure by including all possible intermediate states in the PDG that conserve $\rm J^{PC}$,
where these intermediate states can decay into $\kpkm$, $\pzpz$, $K^\pm\piz$, $\kpkm\piz$ or $K^\pm \pzpz$ final states.
 Then we examine the statistical significance of the individual amplitudes, and drop the ones with statistical significance less than 5$\sigma$.
 The process is repeated until no amplitude remains with a statistical significance less than 5$\sigma$.
After that, all the removed processes are reintroduced individually to make sure that they are not needed in the fit.
 In the above approach, the statistical significance of each individual amplitude is determined by the changes in the negative log likelihood (NLL) value and the number of free parameters in the fit with and without the corresponding amplitude included.

 The above strategy is performed individually on the data sets at $\sqrt{s}= 2.125$ and 2.396~$\gev$, which have the largest luminosities among the ten data sets. 
The nominal solution for data at $\sqrt{s}= 2.125$~$\gev$ includes the two-body decay processes $K^{+}(1460)K^{-}$, $K^{+}_1(1270)K^{-}$, $K^{+}_1(1400)K^{-}$, $\ksks$, $K^{*+}_{0}(1430)K^{*-}(892)$, $\phi(1020) \sigma$, $\phi(1020) f_{0}(980)$, $\phi(1020) f_{2}(1270)$ and $\omega(1420)\pi^{0}$, as well as the three-body decay processes $\kpkm \sigma$, $\kpkm f_{0}(980)$ and $\kpkm f_{0}(1370)$.
 For the data at $\sqrt{s}= 2.396$~$\gev$, the additional intermediate processes $K^{*+}_2(1430) K^{*-}(892)$, $K^{*+}(892)K^-\piz$ and $\phi(1020) f_{0}(1370)$ are included, but without the $\phi(1020)\sigma$ and $\phi(1020)f_2(1270)$ processes.
 An interesting decay mode $K^{*+}(1410)K^{-}$, which is expected to have a sizeable decay rate for a conventional $3 ^3S_1$ $s\bar{s}$ state~\cite{2017ding}, is found to be less than 3$\sigma$ in both data samples.
 In the above, the three-body decays are treated as consecutive quasi two-body decays with a very broad resonance decaying into $\kpkm$ or $K^{+}\pi^{0}$ and modeled as a 1$^{-}$ phase space distribution.
 The intermediate states $K^{+}(1460)$, $K^{+}_1(1270)$, $K^{+}_1(1400)$ decay into $K^{*+}(892)\piz$, and $\omega(1420)$ decays into $K^{*\pm}(892)K^{\mp}$, followed by $K^{*+}(892) \to K^+\piz$.
The state $K^{*+}_{0}(1430)$ decays into $K^+\piz$.
The state $\phi(1020)$ decays into $\kpkm$ and $\sigma$, $f_{0}(980)$, $f_2(1270)$, $f_0(1370)$ decay into $\piz\piz$.
 The masses and widths of the $K(1460)$, $K_1(1400)$, $K_1(1270)$ and $\omega(1420)$ in the fit are determined by scanning the likelihood value, and the results are consistent with the parameters in the PDG.
The masses and widths of other intermediate states are fixed to PDG values. 
The statistical significance of all intermediate processes are summarized in sections II and III of the supplemental material~\cite{supply}, respectively. 
The corresponding comparison of invariant mass spectra and angular distributions between data and MC projections are shown in section IV of the supplemental material.

 For the other eight data samples, due to limited statistics, we do not perform the above optimization strategy to determine which intermediate processes to include. Instead, we use the same intermediate processes as the data sets with nearby c.m. energy.  The data sets with $\sqrt{s}=$ 2.000, 2.100, 2.175, 2.200 and 2.232~$\gev$ (referred to as group I data), use the same processes as $\sqrt{s}=2.125~\gev$, while the other three points (group II data) use the same processes as $\sqrt{s}=2.396~\gev$.

 The total Born cross sections for $\epem\to \kpkm\piz\piz$ and the Born cross sections for the intermediate processes are obtained at each c.m.~energy using:
 \begin{equation}
     \sigma^{B}=\frac{N^{sig}}{\mathcal{L}_{int}~\frac{1}{|1-\Pi|^{2}}~(1+\delta)^{r}~\mathcal{B}r~\epsilon},
     \label{equation}
 \end{equation}
 where $N^{sig}$ is the corresponding signal yield, and is determined by calculating the fraction according to the PWA results for the individual intermediate process;
 $\mathcal{L}_{int}$ is the integrated luminosity;
 $(1+\delta)^{r}$ is the ISR correction factor obtained from a QED calculation~\cite{VR,conexc} and incorporating the input cross section from this analysis iteratively;
 $\frac{1}{|1-\Pi|^{2}}$ is the vacuum polarization factor taken from a QED calculation~\cite{VP};
 $\epsilon$ is the detection efficiency obtained from a PWA-weighted MC sample;
and $\mathcal{B}r$ is the product of branching ratios of the intermediate states as quoted in the PDG~\cite{PDG}. In the decay $\epem \to K^{+}(1460)K^{-}$, the branching fraction of $K(1460) \to \ks\pi$ is included in the measured cross section since it has never been measured.

 Two categories of systematic uncertainties are considered in the measurement of the Born cross sections.
 The first category includes uncertainties associated with the luminosity, track detection, PID, kinematic fit, ISR correction, and the branching fractions of intermediate states.
 The uncertainty associated with the integrated luminosity is 1\% at each energy point~\cite{lum}.
 The uncertainty of the detection efficiency is 1\% for each charged track~\cite{trackerror} and photon~\cite{photonerror}.
 The PID efficiency uncertainty is 1.0\% for each charged track~\cite{trackerror}.
 The uncertainty related to the kinematic fit is estimated by correcting the helix parameters of the simulated charged tracks to match the resolution~\cite{helixsys}.
 The uncertainty associated with the ISR correction factor is estimated to be the difference of $(1+ \delta^{r})\epsilon$ between the last two iterations in the cross section measurement.
 The systematic uncertainties from the branching ratios of intermediate states in the subsequent decays are taken from the PDG~\cite{PDG}.
 The second category of uncertainties are from the PWA fit procedure.
Fits with alternative scenarios are performed, and the changes of signal yields are taken as systematic uncertainties.
 Uncertainties from the BW parameterization are estimated by replacing the constant-width BW with the mass-dependent width. 
Uncertainties associated with the resonance parameters, which are taken from the PDG and fixed in the fit, are estimated by alternative fits superposing additional constraints on these resonance parameters, where the superposed constraints follow Gaussian distributions with widths equal to their uncertainties.
 One thousand fits are performed, and the resultant standard deviations of the signal yields are taken as systematic uncertainties.
 Uncertainties associated with the additional resonances are estimated by alternative fits including the components $K^{*}(1410)K$ or the $K^{*}_{2}(1430)K^{*}$, which are most significant, but less than 5$\sigma$.
 Uncertainties due to the barrier factor are estimated by varying the radius of the centrifugal barrier from 0.7 to 1.0 fm.
 To estimate the uncertainties on the detection efficiency related to the fit parameters in the PWA, one hundred MC samples are generated with five hundred groups of parameters of PWA amplitudes which is sampled from a multi-variable Gaussian function according to their mean values and their covariance error matrix from the nominal fit. The standard deviations of the resultant detection efficiencies are considered as the uncertainties.

 In the above procedure, the uncertainties associated with the barrier factor, resonance parameterization and additional resonances are strongly affected by the statistics. Thus, those uncertainties of data with $\sqrt{s}$=2.125~$\gev$ are assigned to the group I data, while those of data with $\sqrt{s}$=2.396~$\gev$ are assigned to the group II data.
 Assuming all sources of systematic uncertainties are independent, the total uncertainties are the quadratic sums of the individual values, shown in section V of the supplemental material~\cite{supply}, where the sources of the uncertainties tagged with `*' are assumed to be 100\% correlated among each energy points.

 The measured total Born cross sections for $\epem\to\kpkm\piz\piz$ and the Born cross sections for the subprocess $\epem\to\phi\piz\piz$, summing over all the $\piz\piz$ intermediate processes and their interferences, are shown in Fig.~\ref{fig1}. Good agreement is found with the previous results from BaBar.
 In order to study the properties of $1^{--}$ states, the cross sections for the processes $\epem \to K^{+}(1460)K^{-}$,  $K^{+}_{1}(1400)K^{-}$, $K^{+}_{1}(1270)K^{-}$ and $\ksks$, referred to as the $KK$ processes, are shown in Fig.~\ref{09fit}.  A clear peak between 2.1 and 2.2 GeV is present in the process $\epem \to K^{+}(1460)K^{-}$, and dips are observed for the processes $\epem\to K^{+}_{1}(1400)K^{-}$ and $K^{+}_{1}(1270)K^{-}$ in almost the same energy region.
This may be due to destructive interference between different components. No obvious  structure or dip is present in the process $\epem\to \ksks$.
 All the various numbers used in the cross section calculation are summarized in section I of the supplemental material~\cite{supply}.

 \begin{figure}[htbp]
 \centering
\includegraphics[width=0.23\textwidth]{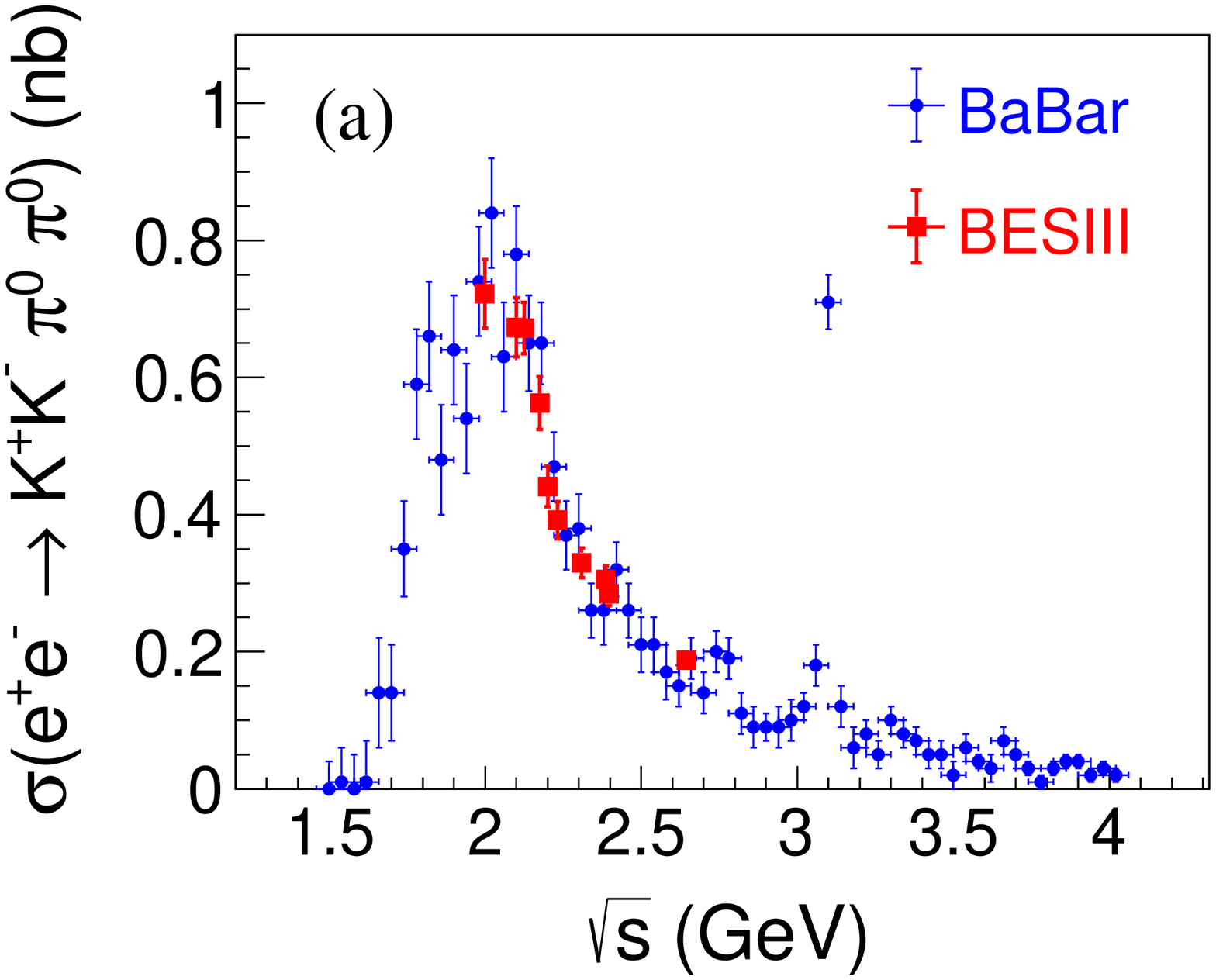}
\includegraphics[width=0.23\textwidth]{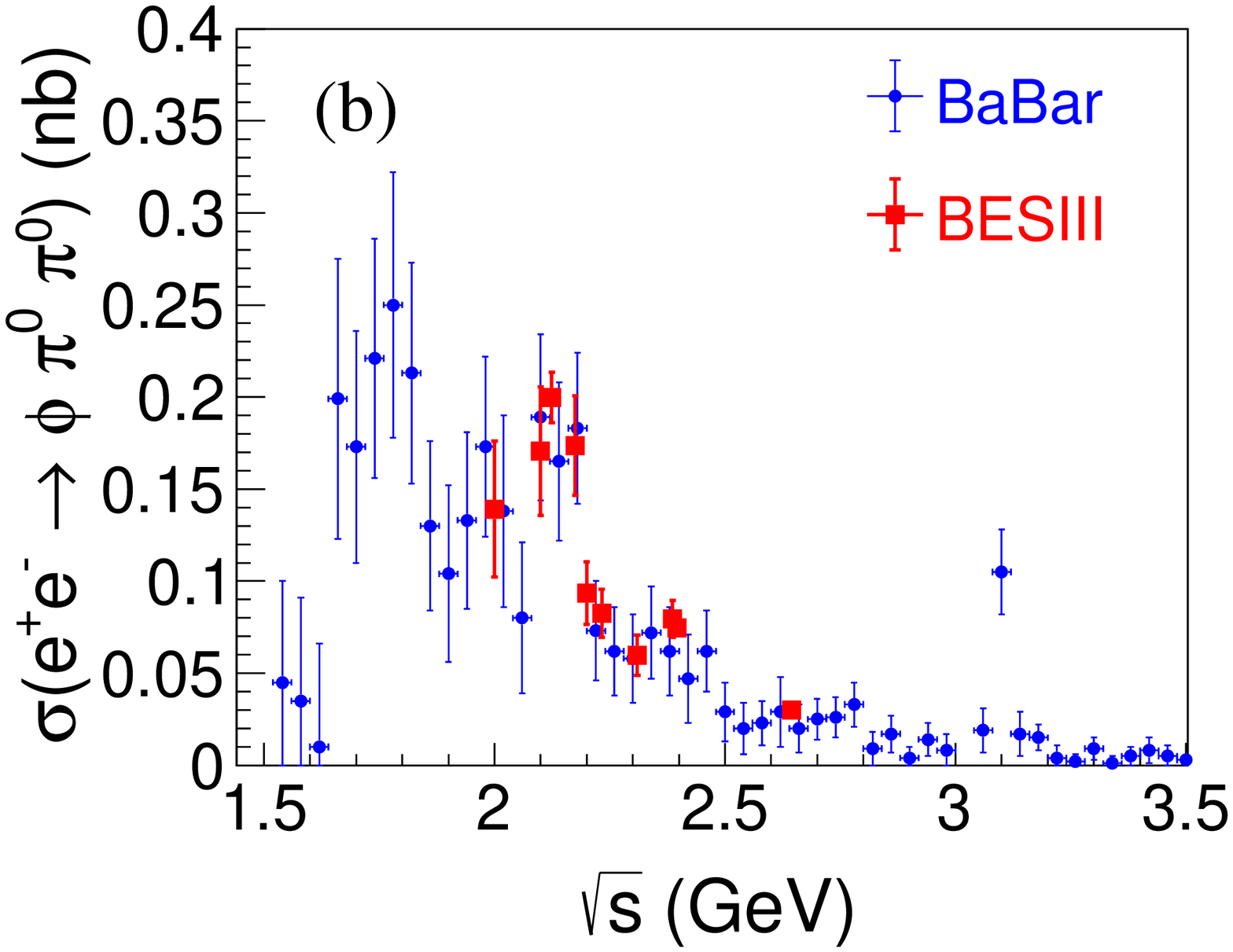}
 \caption{The Born cross sections for (a)~the process $\epem \to \kpkm\piz\piz$ and (b)~the subprocess $\epem \to\phi\piz\piz$.  The red squares are from this analysis; the blue dots are from the BaBar experiment.}
 \label{fig1}
\end{figure}

\begin{figure}[htbp]
 \centering
\includegraphics[width=0.23\textwidth]{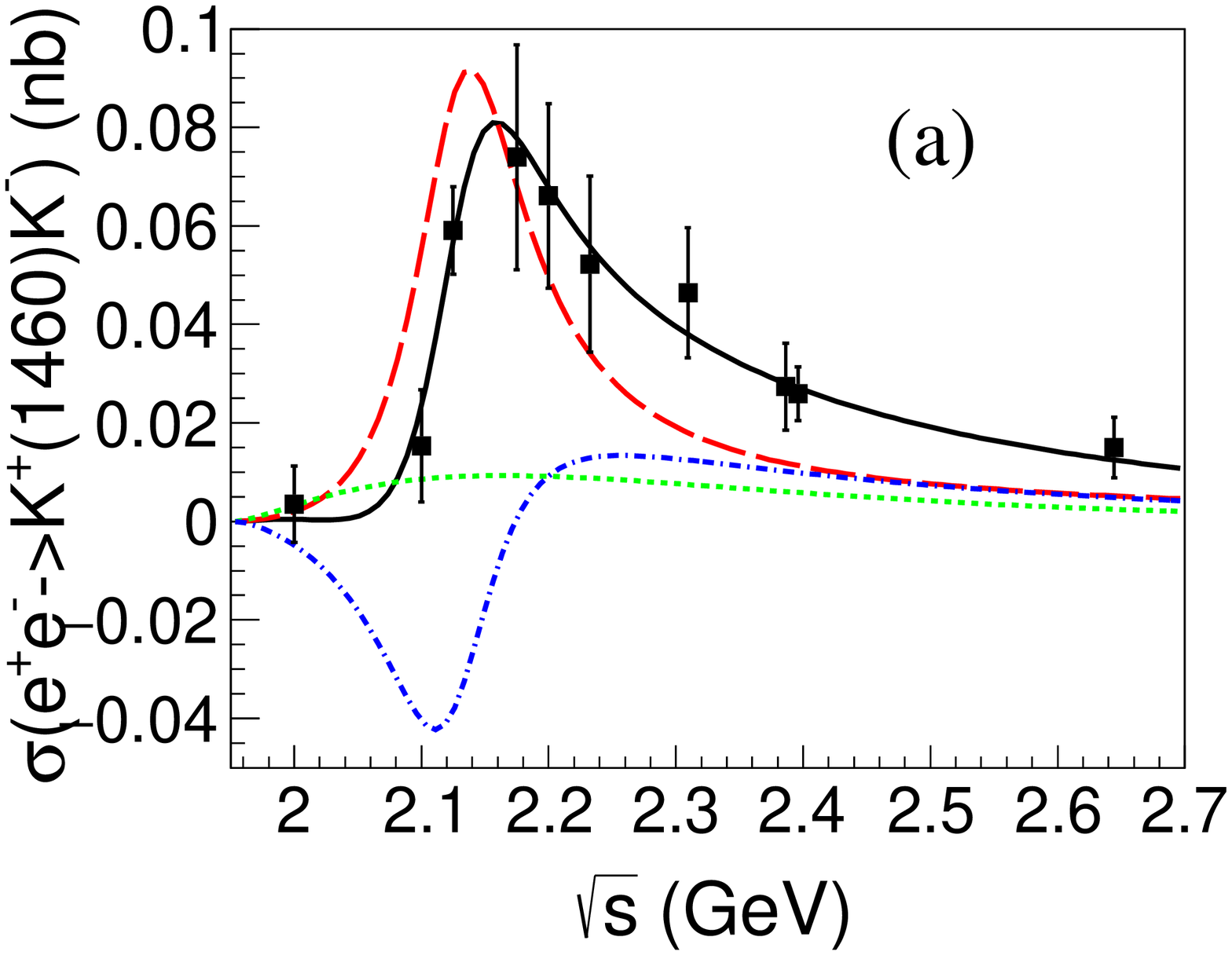}
\includegraphics[width=0.23\textwidth]{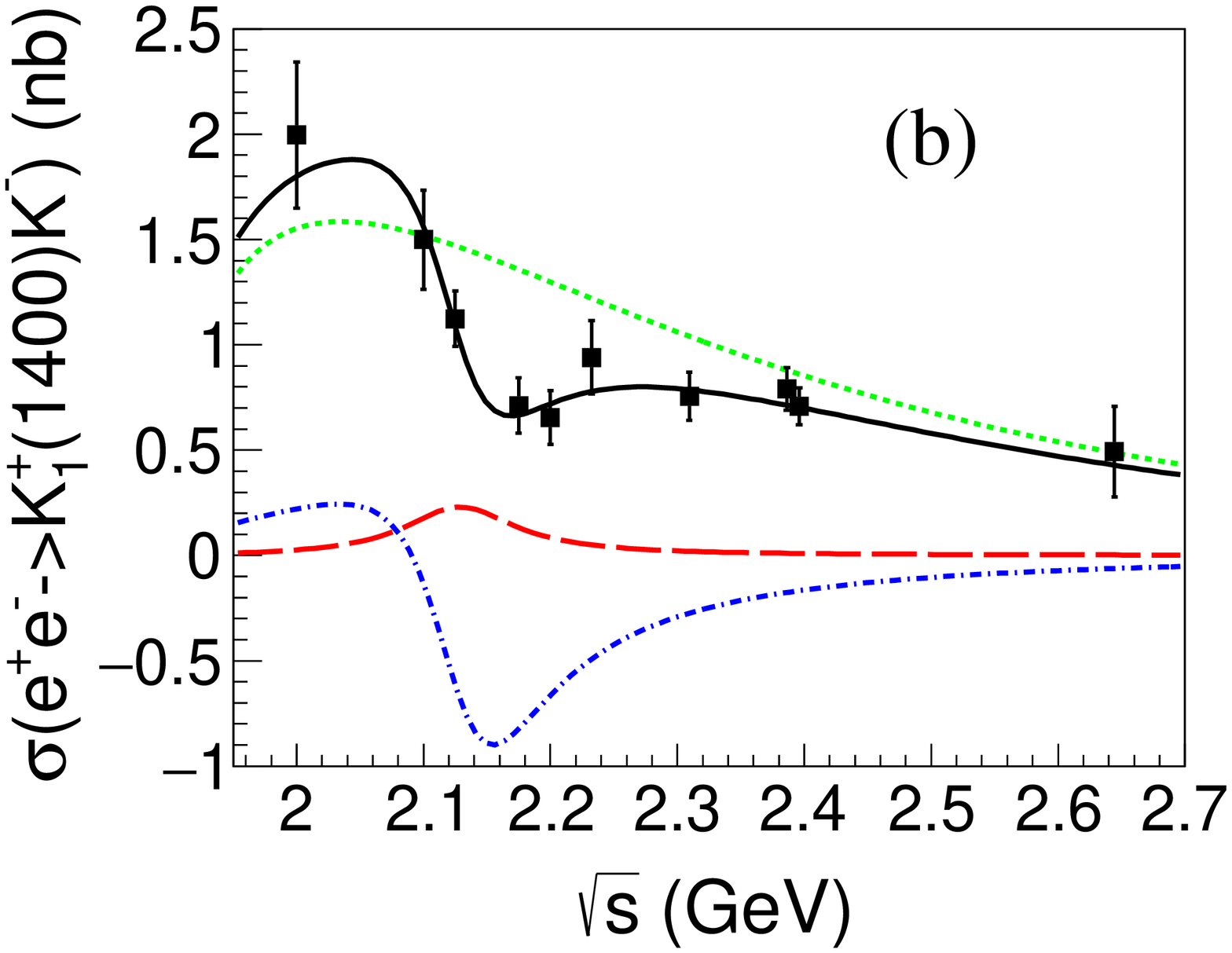}
\includegraphics[width=0.23\textwidth]{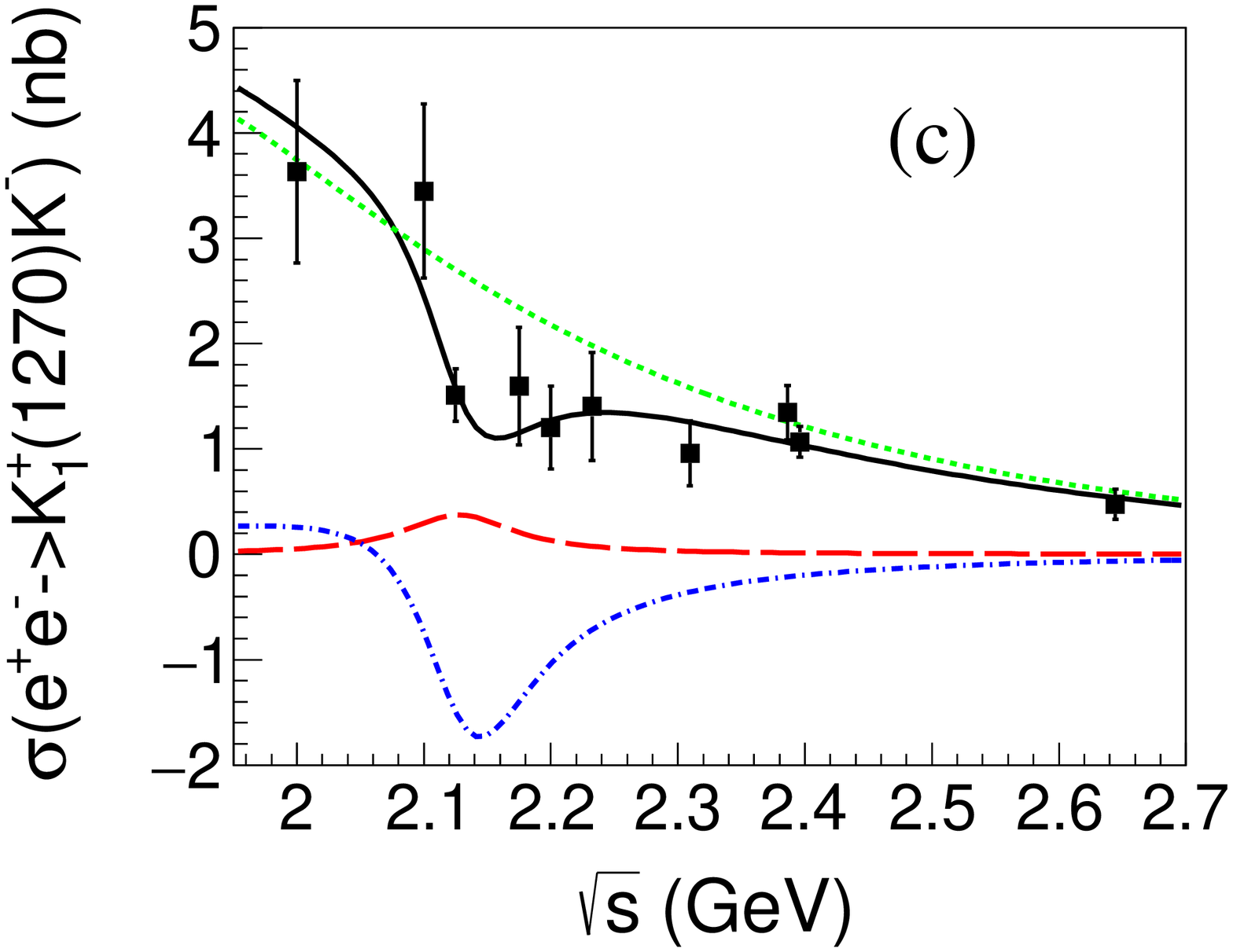}
\includegraphics[width=0.23\textwidth]{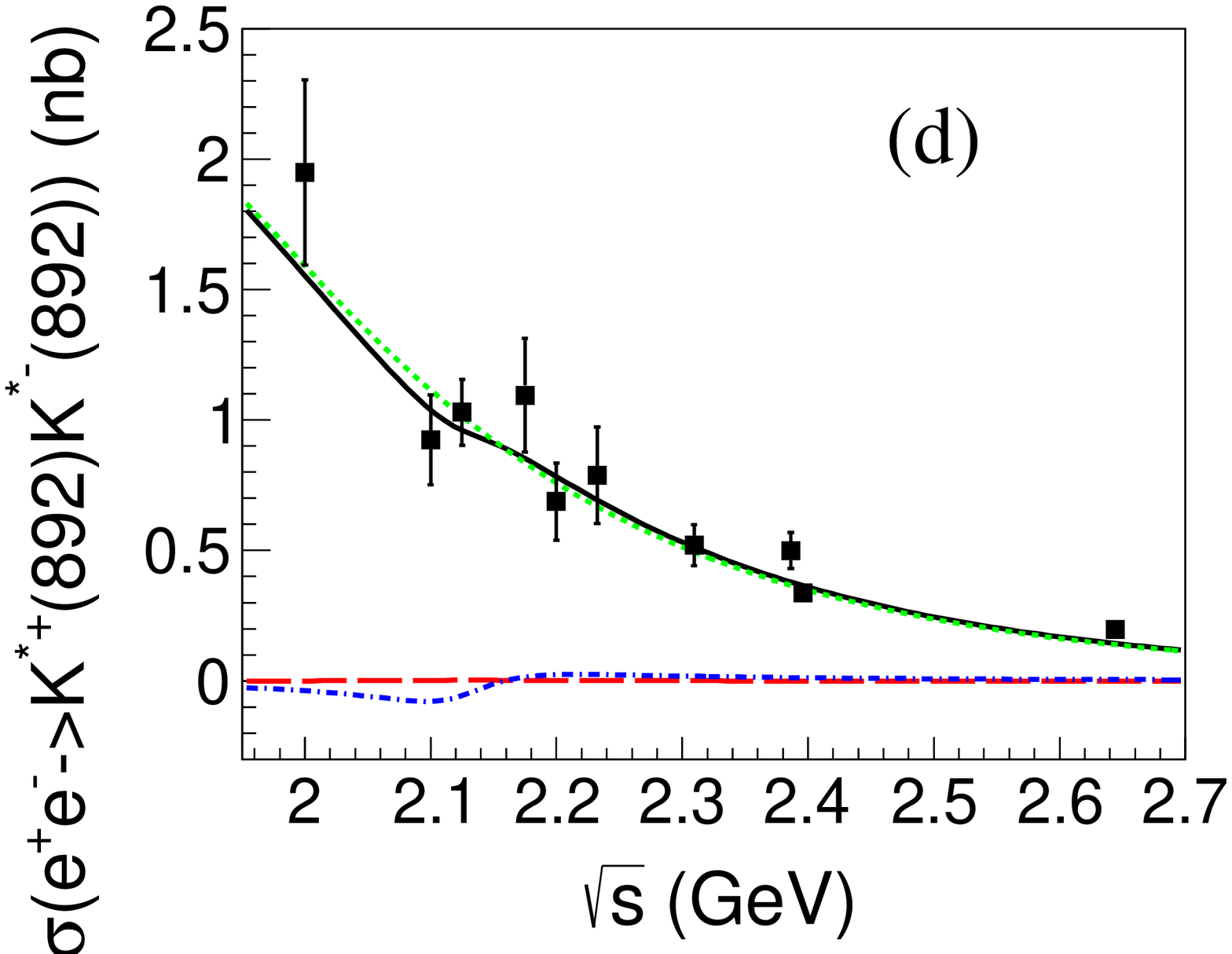}
 \caption{Fit to the cross sections for $e^+e^-$ to the final states (a) $K^{+}(1460)K^{-}$, (b) $K^{+}_{1}(1400)K^{-}$, (c) $K^{+}_{1}(1270)K^{-}$ and (d) $\ksks$, where
 black dots with errors are data, the black solid curves are the overall fit results, the red long-dashed curves are from the intermediate state, the green short-dashed curves are from the continuum component, and the blue dash-dotted curves are the interference contribution for solution 1.}
 \label{09fit}
\end{figure}

 To further examine the structure, a binned $\chi^{2}$ fit, incorporating the correlated and uncorrelated uncertainties among different energy points, is performed to the cross sections for the $K^{+}(1460)K^{-}$, $K^{+}_{1}(1400)K^{-}$, $K^{+}_{1}(1270)K^{-}$ and $\ksks$ processes.
 The fit probability density function (PDF) for the individual processes is the coherent sum of a continuum component $f_1$ and a resonant component $f_2$ :
 \begin{equation}
      \mathcal{A} = f_1+ e^{i\phi} f_2,
 \label{eq1}
 \end{equation}
 where $\phi$ is the relative phase between the two components.
By considering phase space $\Phi(\sqrt{s})$, the energy-dependent cross section of the QED process, and the relative orbital angular momentum $L$ in the two-body decay, the amplitude $f_1$ is described as
\begin{equation}
      f_1 = q^{L}\frac{\sqrt{\Phi(\sqrt{s})}}{s^{n}},
\label{eq10}
\end{equation}
where $q$ is the momentum of the daughter particle.
The resonant amplitude $f_2$ is described with a BW function,
\begin{equation}
      f_2 = \frac{M_R}{\sqrt{s}}\frac{\sqrt{12\pi \mathcal{B}r \Gamma^{e^{+}e^{-}}_R \Gamma_R}}{s-M_R^{2}+iM_R\Gamma_R}(\frac{q}{q_{0}})^{L}\sqrt{\frac{\Phi(\sqrt{s})}{\Phi(M_R)}},
\label{eq2}
\end{equation}
where $M_R$ is the mass of the structure,
$\Gamma_R$ is the total width,
$\Gamma^{\epem}_R $ is its partial width to $\epem$,
$\mathcal{B}r$ is the decay branching fraction to a given final state,
and $q_{0}$ is the momenta of the daughter particle in the rest frame of the parent particle ($M_R$).

 A simultaneous fit, assuming the same structure among the $K^{+}(1460)K^{-}$, $K^{+}_{1}(1400)K^{-}$, $K^{+}_{1}(1270)K^{-}$ and $\ksks$ processes, is performed to the measured cross sections, as shown in Fig.~\ref{09fit}.
 In the fit, $M_R$ and $\Gamma_R$ are shared parameters between the four processes and are floated, while $\it n$, the production $\mathcal{B}r \Gamma^{\epem}_R$, and the relative phase angle $\phi$ are floated and final state dependent.
 For $\epem \to K^{+}_1(1270)K^{-}$ and  $K^{+}_1(1400)K^{-}$, $L=0$, while $L=1$ for the other two modes.
 The fit results have two solutions with equal fit quality, identical $M_R=(2126.5 \pm 16.8)~\mevcc$ and $\Gamma_R=(106.9 \pm 32.1)~\mev$, but different $\mathcal{B}r \Gamma^{\epem}_R$ and $\phi$ for the processes  $\epem \to K^{+}_1(1400)K^{-}$ and $K^{+}_{1}(1270)K^{-}$,
 as summarized in Table~\ref{09fittable}.
 The statistical significance of the structure is estimated with the change of $\chi^2$ ($\Delta\chi^2$) and the number of degrees of freedom ($\Delta$ndof) between the scenarios with and without the structure included in the fit.
 The overall statistical significance is 6.3$\sigma$, obtained with $\Delta\chi^2$=63.8 and $\Delta$ndof=10.
 The significance of the resonant state for each $KK$ process is also estimated and summarized in Table~\ref{09fittable}.
The significances of the resonant state in the processes $\epem\to K^+(1460)K^-$ and $K_1^{+}(1400)K^-$ are greater than 4.5$\sigma$, while no significant signal is found in the other two processes.
 We also estimate the upper limit at the 90\% confidence level on the production $\mathcal{B}r \Gamma^{\epem}_{R}$ to be 1.9~eV for $\epem\to \ksks$ and 12.5(297.6)~eV for $\epem\to K^{+}_{1}(1270)K^{-}$.

\begin{table}[htbp]
\begin{center}
\footnotesize
\caption{A summary of fit results.}
\begin{tabular}{l |c | c |c |c}
\hline\hline
Channel       &   & $\mathcal{B}_{r} \Gamma^{e^{+}e^{-}}_{R}$ (eV) &  $\phi$ (rad)  &  Sig. ($\sigma$) \\
\hline
$K^{+}(1460)K^{-}$                       &            &  3.0 $\pm$ 3.8 &    5.6  $\pm$ 1.5  &   4.4\\
\hline
\multirow{2}{*}{$K^{+}_{1}(1400)K^{-}$}  &  solution 1 &  4.7 $\pm$ 3.3 & 3.7 $\pm$ 0.4  & \multirow{2}{*}{4.8}\\
                                         &  solution 2 &  98.8 $\pm$ 7.8  &   4.5 $\pm$ 0.3   &   \\
\hline
\multirow{2}{*}{$K^{+}_{1}(1270)K^{-}$}  &  solution 1 &  7.6 $\pm$ 3.7 & 4.0 $\pm$ 0.2  & \multirow{2}{*}{1.4}\\
                                         &  solution 2 &  152.6 $\pm$ 14.2    &   4.5 $\pm$ 0.1    &   \\
\hline
$\ksks$                                   &           &0.04 $\pm$ 0.2      &    5.8  $\pm$ 1.9     &   1.2\\
 \hline
  \hline
\end{tabular}
\label{09fittable}
\end{center}
\end{table}

  The systematic uncertainties on the resonant  parameters come from the absolute c.m.\ energy measurement, the measured cross section, and the fit procedure.
  The uncertainty of the c.m.~energy from BEPCII is small, and is ignored in the determination of the parameters of the structure.
  The statistical and systematic uncertainties of the measured cross section are incorporated in the fit, thus no further uncertainty is necessary.
  The uncertainties associated with the fit procedure include those from the fit range and signal model.
  The uncertainty from the fit range is investigated by excluding the last energy point $\sqrt{s} = 2.644$ GeV in the fit. The resultant changes, 5.1 $\mevcc$ for mass and 9.1 MeV for width, are taken as the systematic uncertainties.
  To assess the systematic uncertainty associated with the signal model, an alternative BW function with energy-dependent width is implemented in the fit, and results in differences of 11.3 $\mevcc$ and 26.5~$\mev$ for mass and width, respectively, which are taken as the systematic uncertainties.
  The overall systematic uncertainties are the quadratic sum of the individual ones, 12.4~$\mevcc$ and 28.1~$\mev$ for the mass and width, respectively.

In summary, a PWA of the process $\epem \to \kpkm\pzpz$ is performed for ten data samples with c.m.~energies from 2.000 to 2.644~$\gev$ and with an integrated luminosity of 300~$\ipb$.
The Born cross sections for $\epem\to\kpkm\pzpz$ and $\phi\pzpz$ are obtained and are consistent with those from the BaBar experiment.
We also measure the cross sections for the processes $\epem \to K^{+}(1460)K^{-}$, $K^{+}_{1}(1400)K^{-}$, $K^{+}_{1}(1270)K^{-}$, and $\ksks$, individually, and perform a simultaneous fit on the obtained results.
The fit results in a structure with mass $M$ = (2126.5 $\pm$ 16.8 $\pm$ 12.4)~$\mevcc$, width $\Gamma$ = (106.9 $\pm$ 32.1 $\pm$ 28.1)~$\mev$, and statistical significance 6.3~$\sigma$, where the uncertainties are statistical and systematic, respectively.
The structure is directly produced in $\epem$ collisions, thus has $J^{PC}=1^{--}$.
This structure has a mass close to the masses of the vector particles $\phi(2170)$, $\rho(2150)$ and $\omega(2290)$ listed in the PDG~\cite{PDG}.
Its width is only consistent with the $\phi(21770)$ and is different from the others by more than 3$\sigma$.

Assuming the observed structure is $\phi(2170)$, our measurement implies that the $\phi(2170)$ has a sizable partial width to $K^+(1460)K^-$, $K_1^+(1400)K^-$, and $K_1^+(1270)K^-$, but a much smaller partial width to $K^{*+}(892)K^{*-}(892)$ and $K^{*+}(1410)K^-$. 
According to Ref.~\cite{2017ding}, the $3 ^3S_1$ $s\bar{s}$ state mainly decays to $K^{*+}(892)K^{*-}(892)$ and $K^{*+}(1410)K^-$, but has a much smaller partial width to $K_{1}^+(1400)K^-$ and $K^{+}(1460)K^{-}$.
On the other hand, the $2 ^3D_1$ $s\bar{s}$ state has an expected partial width to $K_{1}^+(1400)K^-$ smaller than that to $K^{*+}(1410)K^-$ by a factor of 2-5~\cite{2017ding,2017wang}.
A hybrid state is expected to decay dominantly into $K_1^+(1270)K^-$ and $K_1^+(1400)K^-$, while it should be highly suppressed in the modes $K^{*+}(892)K^{*-}(892)$ and $K^+(1460)K^-$~\cite{2017ding2}. 
None of the above theoretical expectations are in good agreement with our experimental results.

The BESIII collaboration thanks the staff of BEPCII and the IHEP computing center for their strong support. This work is supported in part by National Key Basic Research Program of China under Contract No. 2015CB856700; National Natural Science Foundation of China (NSFC) under Contracts Nos. 11625523, 11635010, 11735014, 11822506, 11835012; the Chinese Academy of Sciences (CAS) Large-Scale Scientific Facility Program; Joint Large-Scale Scientific Facility Funds of the NSFC and CAS under Contracts Nos. U1532257, U1532258, U1732263, U1832207; CAS Key Research Program of Frontier Sciences under Contracts Nos. QYZDJ-SSW-SLH003, QYZDJ-SSW-SLH040; 100 Talents Program of CAS; INPAC and Shanghai Key Laboratory for Particle Physics and Cosmology; ERC under Contract No. 758462; German Research Foundation DFG under Contracts Nos. Collaborative Research Center CRC 1044, FOR 2359; Istituto Nazionale di Fisica Nucleare, Italy; Ministry of Development of Turkey under Contract No. DPT2006K-120470; National Science and Technology fund; STFC (United Kingdom); The Knut and Alice Wallenberg Foundation (Sweden) under Contract No. 2016.0157; The Royal Society, UK under Contracts Nos. DH140054, DH160214; The Swedish Research Council; U. S. Department of Energy under Contracts Nos. DE-FG02-05ER41374, DE-SC-0010118, DE-SC-0012069; University of Groningen (RuG) and the Helmholtzzentrum fuer Schwerionenforschung GmbH (GSI), Darmstadt.

\end{document}